# The origin of preferential attachment and the generalized preferential attachment for weighted networks

Chen Chen[1]

Department of Physics, Sun Yat-sen University, Guangzhou 510275, People's Republic of China

**Abstract**: In this paper, we first discuss the origin of preferential attachment. Then we establish the generalized preferential attachment which has two new properties; first, it encapsulates both the topological and weight aspects of a network, which makes it is neither entirely degree preferential nor entirely weight preferential. Second, it can tell us not only the chance that each already-existing vertex being connected but also how much weight each new edge has. The generalized preferential attachment can generate four power-law distributions, besides the three for vertex degrees, vertex strengths, and edge weights, it yield a new power-law distribution for the subgraph degrees.

Many systems in the world can be described as complex networks, which are structures of vertices and edges. For examples, the protein networks [1], the WWW web [2], and the scientific collaboration networks [3,4]. In the past few years, complex networks have been widely studied by scientists form various areas [5-7], and many properties of complex networks, such as the small-world character [8] and the scale-free behavior [9], have been revealed. The small-world networks, introduced by Watts and Strogatz, are a kind of networks between the two extreme cases; regular lattices and random graphs. Specifically, they have high clustering coefficients like regular lattices, yet have short average path lengths like random graphs. Besides the small-world character, the scale-free behavior is another property, which refers to the fact that many large networks show power-law degree distributions $P(k) \sim k^{-\gamma}$, where $P(k)$ is defined as the probability that a randomly selected vertex has exactly $k$ edges. And it is believed that the degree preferential attachment mechanism, introduced by Barabási and Albert, is an effective method of generating the scale-free property [5].

---

[1]Corresponding author.
email address: st03chen@hotmail.com, st03chc@student.sysu.edu.cn

This degree preferential attachment (DPA) was also applied to the study of weighted networks [10-12], in which the strength of the interactions between vertices is considered. And subsequently, the weight preferential attachment (WPA) was brought forward [10,13,14]. These two kinds of preferential attachments focus on different sorts of network properties. The DPA focus on the *degree* or the *topological* aspect of a network, defined as

$$\prod_{new \to i} = \frac{k_i}{\sum_j k_j}. \tag{1}$$

Namely, the probability $\prod_{new \to i}$ of an already-existing vertex $i$ being chosen for connecting is proportional to its degree $k_i$. On the other hand, WPA focus on the *weight* aspect, defined as

$$\prod_{new \to i} = \frac{s_i}{\sum_j s_j}. \tag{2}$$

That is, the probability $\prod_{new \to i}$ of an already-existing vertex $i$ being chosen for connecting is proportional to its strength $s_i$. The strength $s_i$ of vertex $i$ is defined as $s_i = \sum_{j \in N(i)} w_{ij}$, where $j$ runs over the neighbors $N(i)$ of vertex $i$ [10] and $w_{ij}$ is the weight of the edge connecting vertices $i$ and $j$, which characterize the interaction strength between $i$ and $j$. Typical definition of $w_{ij}$ can be found in Ref. [3,4,15]. In this paper, we will consider only undirected cases, where the weights are symmetric ($w_{ij} = w_{ji}$).

These two kinds of preferential attachment could be regarded as two extreme cases. And when scientists modeling weighted evolving networks, they often chose one of them, either the DPA [10-12] or the WPA [10,13,14]. However, just as real-world networks are neither completely regular nor completely random, real-world preferential attachment should be neither entirely degree preferential nor entirely weight preferential. We should consult both topological aspect and weight aspect into one preferential attachment. Besides, these two kinds of preferential attachment can only tell us which vertices are likely to be selected. They can't tell us what weights the new edges are. Thus, when scientists study the weighted evolving networks, they often use other mechanism to assign weight. For examples, when an already-existing vertex $i$ is chosen for connecting by a new edge, the weight of the new edge is determined by the degree of $i$ [10], or fixed to a constant when they first appear (but will change later)[13]. We argue that a preferential attachment used in weighted evolving network models should tell us not only which vertices are likely to be selected, but also what weights the new edges are. In the following part of this paper, we will first survey the origin of preferential attachment, and then we will raise the generalized preferential attachments (GPA) which satisfy these two requirements. And finally, we will study the statistical properties of the generated networks.

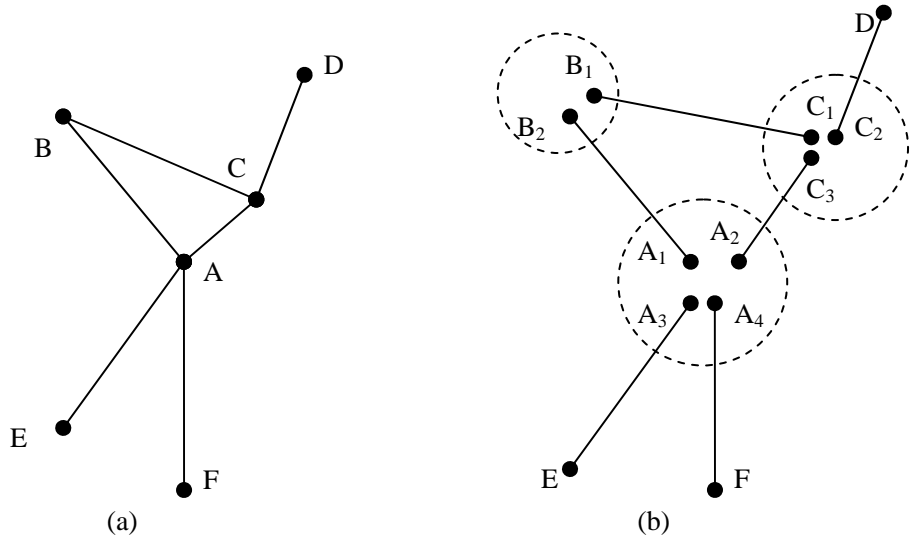

Fig. 1. A vertex $i$ in (a) with degree $k_i$ can be treated as a crowd of $k_i$ vertices, as seen in (b), the circle means that the vertices inside belong to the same crowd, and each vertex in crowd has only one edge.

When we use the DPA, two vertices are equivalent, that is, have the same chance of being chosen for connecting, when they have the same degree. But when we use the WPA, two vertices are equivalent iff they have the same strength. This concept, the equivalence of two vertices, plays a key role in the evolving of a network. On the one hand, as mentioned above, each preferential attachment corresponds to one kind of vertex equivalence. On the other hand, this concept is more essential. From a definition of vertex equivalence and some basic hypothesis of statistical mechanics, we can deduce a preferential attachment. Specifically, as shown in Fig. 1(a), suppose in this network, two vertices are equivalent when they have the same degree. We are now going to choose one from the six already-existing vertices, so what is the chance of each vertex being selected? Assume the network is a non-interaction system, in which vertices keep their probabilities of being selected no matter how *near* they are (even two vertices overlap). Due to this assumption, a vertex $i$ with $k_i$ edges can be considered as the overlapping of a crowd of vertices $i_j$ ( $j = 1, 2, \cdots, k_i$ ), in which each vertex has only one edge, as seen in Figs. 1 (b). Thus, vertex $i$ has the same chance of being selected as the corresponding vertex crowd; here a vertex crowd being selected means any of its vertex $i_j$ ( $j = 1, 2, \cdots, k_i$ ) being selected. Since we've define "two vertices equivalent when they have the same degree", all vertices in Figs. 1(b) have equal chances to be selected, because each of them has only one edge. Therefore, the probability $\prod_{new \to i}$ of vertex $i$ being chosen is proportional to the number of vertices in the corresponding crowd, namely, $\prod_{new \to i} \propto k_i$. So we get DPA from "two vertices are equivalent when they have the same degree". Analogously, if two vertices are equivalent when they have the same strength, we will get WPA, $\prod_{new \to i} \propto s_i$. We can also define two vertices equivalent when they have both the same degree and weight, this time the preferential attachment will be $\prod_{new \to i} \propto a_1 k_1 + a_2 s_2$, where $a_1$ and $a_2$ are the coefficients of proportionality which evaluate the importance of degrees and strengths, respectively.

Since we want a preferential attachment to include both the topological aspect and weight aspects, naturally we can define two vertices equivalent when they have the same degree and weight But these two values, degree and strength, are insufficient to describe a vertex comprehensively in a weighted network. Two vertices with the same degree and strength can still be quite different. We can easily tell a vertex with eight edges whose edge weight is 6, from another vertex who has also eight edges but four edges with weight 2, while the other four with weight 10, thought these two vertices have the same degree and weight. To solve this problem, we introduce the degree distribution of a vertex (it is different from the degree distribution of a network). The degree distribution of a vertex $i$ points out the number of edges connecting to $i$ with edge weight between $w$ and $w + \Delta w$. Specifically, we divide the range of edge weight into several intervals $[w_0, w_1) \cup [w_1, w_2) \cup \cdots \cup [w_{n-1}, w_n)$ , and then use a vector $\mathbf{k}_i = (k_{i1}, k_{i2}, \cdots, k_{in})^T$ to represent the degree distribution of vertex $i$ , where $k_{ij}$ equals to the number of edges connecting to $i$ whose edge weight belong to interval $[w_{j-1}, w_j)$ . Apparently, vector $\mathbf{k}_i = (k_{i1}, k_{i2}, \cdots, k_{in})^T$ satisfy

$$k_i = \sum_{j=1}^n k_{ij} \,. \tag{3}$$

where $k_i$ is the degree of vertex $i$ . The division of edge-weight range is depending on the precision we need. And once we choose a division $[w_0, w_1) \cup \cdots \cup [w_{n-1}, w_n)$ , edges with weight in the same interval (e.g. $[w_{m-1}, w_m)$ ) will be considered indiscriminable, thus we can use $\overline{w}_m = (w_{m-1} + w_m) / 2$ instead of their original weight. The division of the edge-weight range is also a division of the network. Once we choose a division $[w_0, w_1) \cup \cdots \cup [w_{n-1}, w_n)$ , the primal network is also divided into $n$ subgraphs and each subgraph corresponds to one interval of the edge-weight range division. In the $j$ th subgraph, all vertices are contained, but only the edges whose weights belong to interval $[w_{j-1}, w_j)$ can be included. Thus, this is substantially dividing a weighted network into several unweighted subgraphs. And the elements in degree distribution vector $k_{i1}, k_{i2}, \cdots, k_{in}$ are just the degrees of vertex $i$ in these subgraphs.

To raise the GPA which satisfied the requirement mentioned above. We define two vertices equivalent when they have the same degree distributions. Then the GPA can be written as

$$\prod_{new \to i} \propto a_1 k_{i1} + a_2 k_{i2} + \cdots + a_n k_{in} = (a_1, a_2, \cdots, a_n) \begin{pmatrix} k_{i1} \\ k_{i2} \\ \vdots \\ k_{in} \end{pmatrix} = \mathbf{a}^T \mathbf{k}_i \,. \tag{4}$$

where $\mathbf{a} = (a_1, a_2, \cdots, a_n)^T$ and $a_j$ reflects the importance of the $j$ th subgraph. To be specific, when all subgraphs are the same important, namely, $a_j = 1$ for $j = 1, 2, \cdots, n$ . From equation (4) we get DPA, $\prod_{new \to i} \propto k_{i1} + k_{i2} + \cdots + k_{in} = k_i$ . And when the importance of a subgraph is proportional to its average edge weight $\bar{w}_j$, that is $a_j = \bar{w}_j$, we then get the WPA, $\prod_{new \to i} \propto \bar{w}_1 k_{i1} + \bar{w}_2 k_{i2} + \cdots + \bar{w}_n k_{in} = s_i$. But in other cases, such as $a_j = \sin \bar{w}_j$, $a_j = \bar{w}_j^2$, or $a_j = 1/\bar{w}_j$, we will get a preferential attachment with both degree and weight aspects included, however, it is neither entirely degree preferential nor entirely weight preferential. Equally important, we hope the GPA can tell us how much weight each new edge is going to have. Remember that we archive our first goal by using a degree distribution vector $\mathbf{k}_i = (k_{i1}, k_{i2}, \cdots, k_{in})^T$ instead of the degree $k_i$. Our second goal can be realize by replacing the probability $\prod_{new \to i}$ by a probability vector $\prod_{new \to i} = (\prod_{new \to i1}, \prod_{new \to i2}, \cdots, \prod_{new \to in})^T$. Here $\prod_{new \to ij}$ represent the probability that vertex $i$ being selected and the new edge weight is $\bar{w}_j$, and obviously, they satisfy $\prod_{new \to i} = \sum_j \prod_{new \to ij}$. Then, we generalized equation (4) into

$$\prod_{new \to i} \propto \mathbf{A} \mathbf{k}_i \qquad (5)$$

where $\mathbf{A} = \{a_{jl}\}$ is the coefficient matrix, in which $a_{jl}$ evaluates how much $k_{il}$ contribute to the chance that vertex $i$ get a new edge and the edge weight is $\bar{w}_j$. And they satisfy $a_l = \sum_{j=1}^n a_{jl}$, where $a_l$ is the coefficient in formula (4). In addition, analogously to Ref. [16, 17], there could be a nonzero probability that a new vertex attaches to an isolated vertex, i.e.

$$\prod_{new \to i} \propto \mathbf{A}(\mathbf{k}_i + \mathbf{b}). \qquad (6)$$

where $\mathbf{b} = (b_1, b_2, \cdots, b_n)^T$ is the initial attractiveness vector in which $b_j$ is the initial attractiveness of the $j$ th subgraph. Now we write out the complete form of the generalized preferential attachment which satisfies all our goals,

$$\prod_{new \to i} = \frac{1}{\rho} \mathbf{A}(\mathbf{k}_i + \mathbf{b}). \qquad (7)$$

where $\prod_{new \to i} = (\prod_{new \to i1}, \prod_{new \to i2}, \cdots, \prod_{new \to in})^T$, $\rho = \sum_{i,q,\eta} a_{\eta q}(k_{iq} + b_q)$, and $\mathbf{A} = (a_{ij})$ is the coefficient matrix.

In the following we will study one simple but useful case that $a_{jl}$ can be described as a product of two items,

$$a_{jl} = c_j d_l. \tag{8}$$

Remember that $a_{jl}$ is a coefficient that evaluates how much the $l$th subgraph contributes to the chance that vertex $i$ gets a new edge and the edge weight is $\overline{w}_j$. Equation (8) separates these two aspects: how much the $l$th subgraph contributes to the chance that a vertex being selected, which is proportional to $d_l$, and the likelihood that the new edge weight is $\overline{w}_j$, which is proportional to $c_j$. We can see the latter aspect much clearer if we substituting equation (8) into (7), which gives

$$\frac{\prod_{new \to iq}}{\prod_{new \to i}} = \frac{\prod_{new \to iq}}{\sum_{\eta} \prod_{new \to i\eta}} = \frac{c_q}{\sum_{\eta} c_{\eta}}. \tag{9}$$

We will use this equation in the following part.

We can solve this case analytically by continuous approximation [5]. Assume that the division of the edge-weight range is $[w_0, w_1) \cup [w_1, w_2) \cup \cdots \cup [w_{n-1}, w_n)$, We start from a random graph ($m_0$ vertices) and each edge being given a weight randomly selected from $\{\overline{w}_1, \overline{w}_2, \cdots \overline{w}_n\}$ (where $\overline{w}_j$ is the median value of interval $[w_{j-1}, w_j)$, namely, $\overline{w}_j = (w_{j-1} + w_j)/2$). And each step, we add a new vertex with $m$ edges that link the new node to $m$ different already-existing vertices. When choosing the $m$ different already-existing vertices and the weight of the $m$ new edges, we use the preferential attachment given by equation (7), in which **A** satisfies equation (8). The time is measured with respect to the number of vertices added to the network, defined as $t = N - m_0$, where $N$ is the size of the network.. The average degree distribution $\mathbf{k}_i(t)$ of vertex $i$ at time $t$ satisfies

$$\frac{d\mathbf{k}_i}{dt} = m\frac{1}{\rho}\mathbf{A}(\mathbf{k}_i + \mathbf{b}) = m\frac{1}{\rho}\mathbf{c}\mathbf{d}^T(\mathbf{k}_i + \mathbf{b}) \tag{10}$$

where $\mathbf{c} = (c_1, c_2, \cdots, c_n)^T$, $\mathbf{d} = (d_1, d_2, \cdots, d_n)^T$ and $\rho$ is given by

$$\rho = \sum_{i,\eta,q} a_{\eta q}(k_{iq} + b_q) = \sum_{i,\eta,q} c_\eta d_q(k_{iq} + b_q)$$

$$= \left(\sum_\eta c_\eta\right)\left(\sum_i \sum_q (k_{iq} + b_q)\right) = \left(\sum_\eta c_\eta\right)\left(\sum_q \left(d_q \sum_i k_{iq}\right) + t\sum_q d_q b_q\right). \tag{11}$$

According to equation (9), we have

$$\frac{\sum_i k_{iq}}{\sum_i k_i} = \frac{\Pi_{new \to iq}}{\sum_\eta \Pi_{new \to i\eta}} = \frac{c_q}{\sum_\eta c_\eta} \tag{12}$$

where $\sum_i k_i = 2mt$. Substituting into (11), we obtain

$$\rho = \left[2m\sum_q d_q c_q + \left(\sum_\eta c_\eta\right)\left(\sum_q d_q b_q\right)\right]t. \tag{13}$$

Solve equation (10) with (13) and the initial condition $\mathbf{k}_i(t_i) = m\mathbf{c}/\sum_\eta c_\eta$, we get

$$\mathbf{k}_i(t) = \frac{m\mathbf{c}}{\sum_\eta c_\eta}\left(\frac{t}{t_i}\right)^\beta \tag{14}$$

where

$$\beta = \frac{m\sum_q d_q c_q}{2m\sum_q d_q c_q + \left(\sum_\eta c_\eta\right)\left(\sum_q d_q b_q\right)} = \frac{m\mathbf{c}'\mathbf{d}^T}{2m\mathbf{c}'\mathbf{d}^T + \mathbf{b}\mathbf{d}^T} \tag{15}$$

and $\mathbf{c}' = \mathbf{c}/\sum_\eta c_\eta$.

Now let's see the properties of the generated network. Equation (14) indicates all subgraphs behave as $P_j(k) \sim k^{-\gamma}$ as $t \to \infty$, where $P_j(k)$ is the degree distribution of the $j$th subgraph. And $\gamma$ is given by [5]

$$\gamma = \frac{1}{\beta} + 1 = 3 + \frac{\mathbf{b}\mathbf{d}^T}{m\mathbf{c}'\mathbf{d}^T}. \tag{16}$$

The average total degree $k_i(t)$ and strength $s_i(t)$ of vertex $i$ at time $t$ also evolve in power-law ways

$$k_i(t) = \sum_l k_{il} = m \left(\frac{t}{t_i}\right)^\beta,$$  (17)

$$s_i(t) = \sum_l \overline{w}_l k_{il} = \left(\sum_l \overline{w}_l c_l\right) \frac{m}{\sum_\eta c_\eta} \left(\frac{t}{t_i}\right)^\beta$$  (18)

which suggest as $t \to \infty$, the total degree distribution and the strength distribution also follow power laws, i.e., $P(k) \sim k^{-\gamma}$, $P(s) \sim s^{-\gamma}$, respectively, where $\gamma$ is given by equation (16) too. The distribution of single edge weight $P(w)$ can be obtained from equation (12), since $P(w = \overline{w}_q) = \sum_i k_{iq} / \sum_i k_i$, for $q = 1, 2, \cdots, n$, we have

$$P(w = \overline{w}_q) = \frac{\sum_i k_{iq}}{\sum_i k_i} = \frac{\prod_{new \to iq}}{\sum_\eta \prod_{new \to i\eta}} = \frac{c_q}{\sum_\eta c_\eta}.$$  (19)

Namely, $P(w)$ is determined only by $\mathbf{c}$.

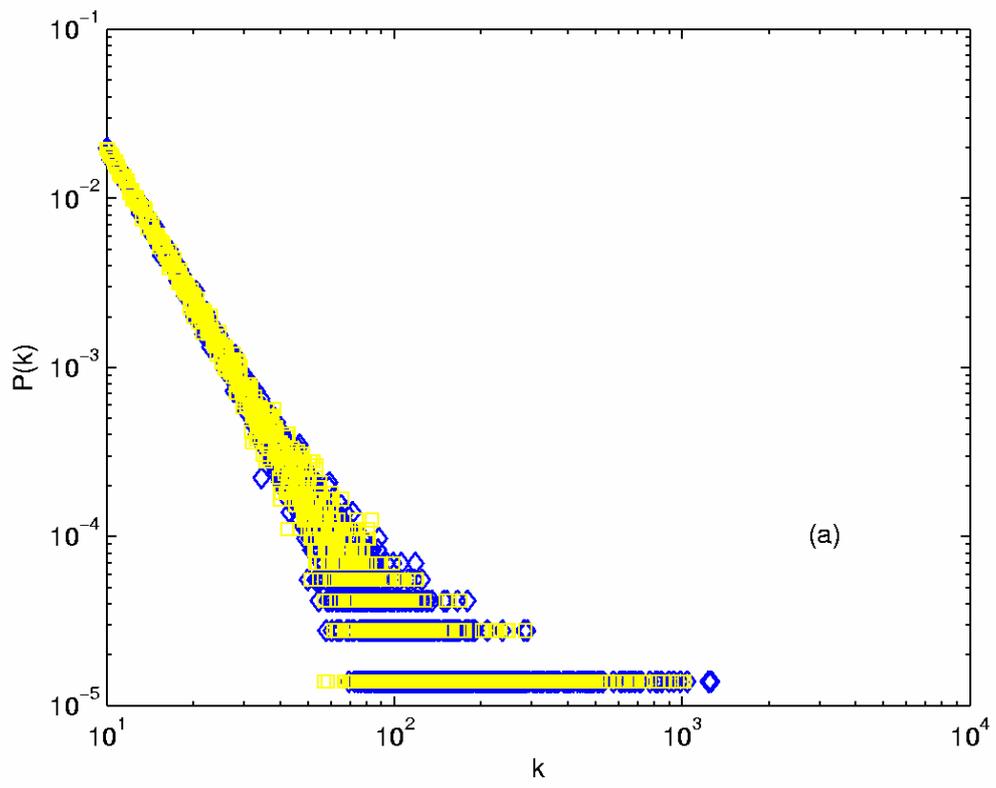

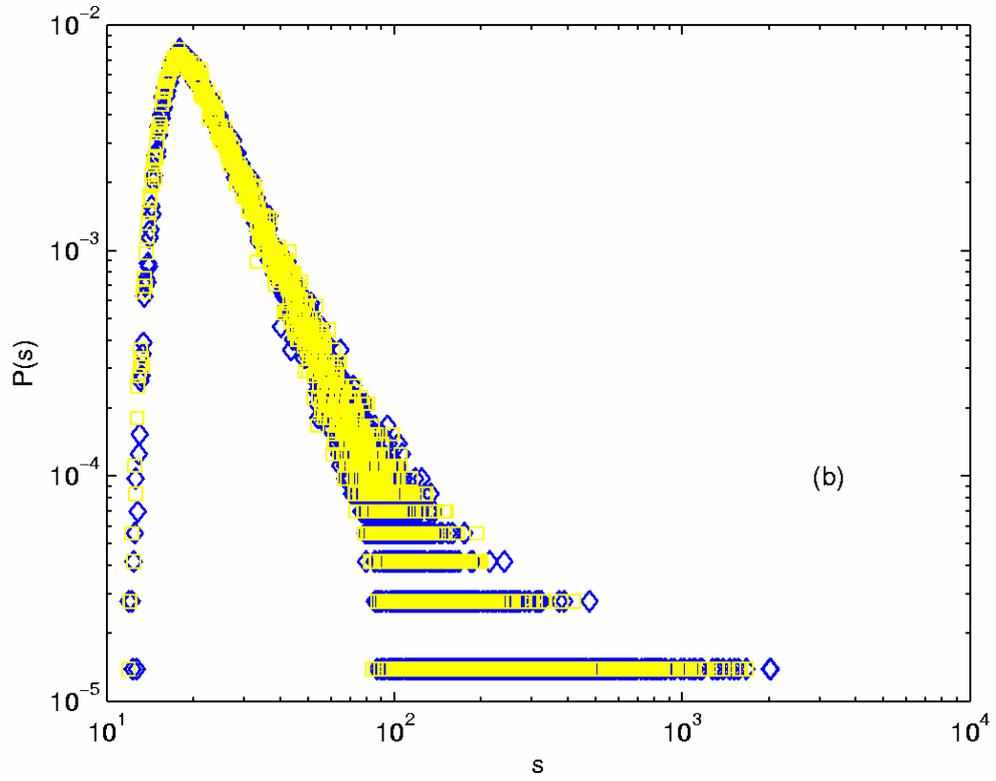

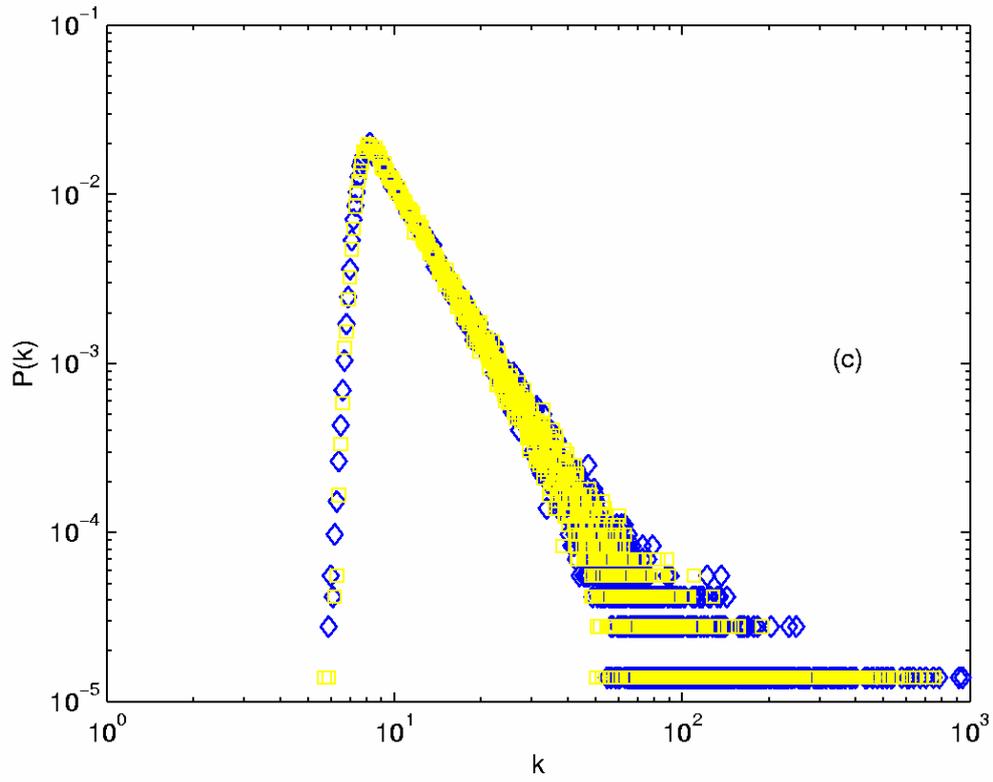

FIG.2 Numerical results in which $b = 0$, $n = 20$, $m = 10$ and $\zeta = 2.5$. (a) Distribution $P(k)$ of total degree. The symbols correspond to different value of $p$, i.e., $p = 1$ ($\diamond$), $20$ ($\square$). (b) Distribution $P(w)$ of strength. (c) Distribution $P(k)$ of the first subgraph. The data are averaged over 10 independent runs of size $N = 72000$.

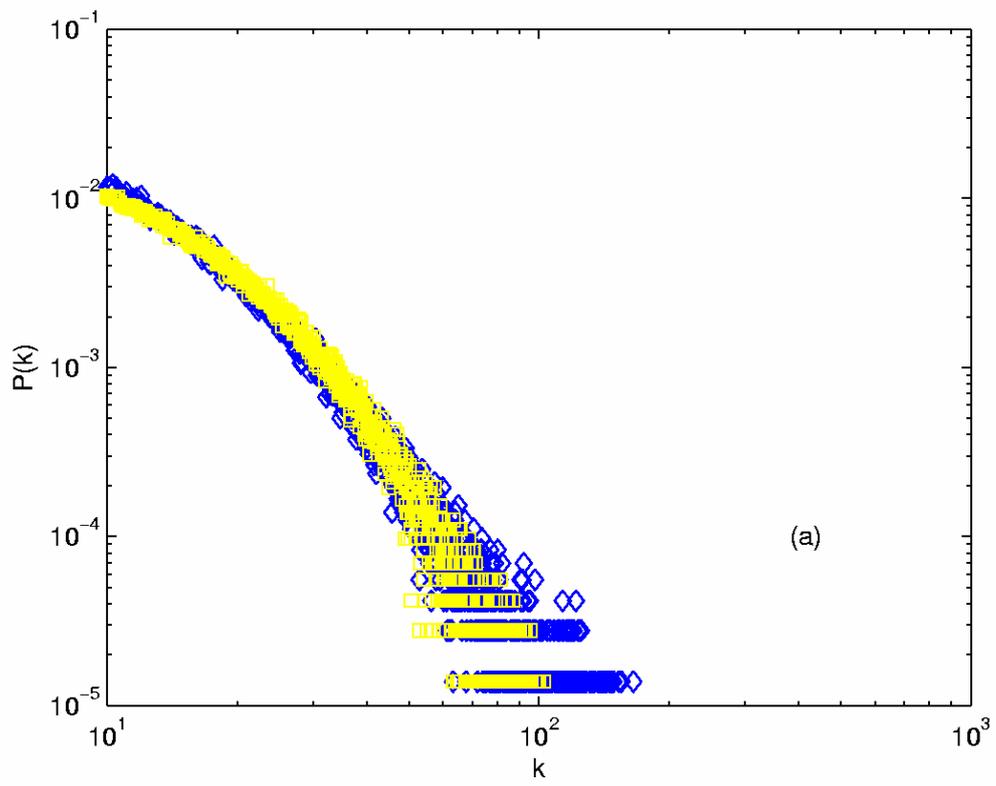

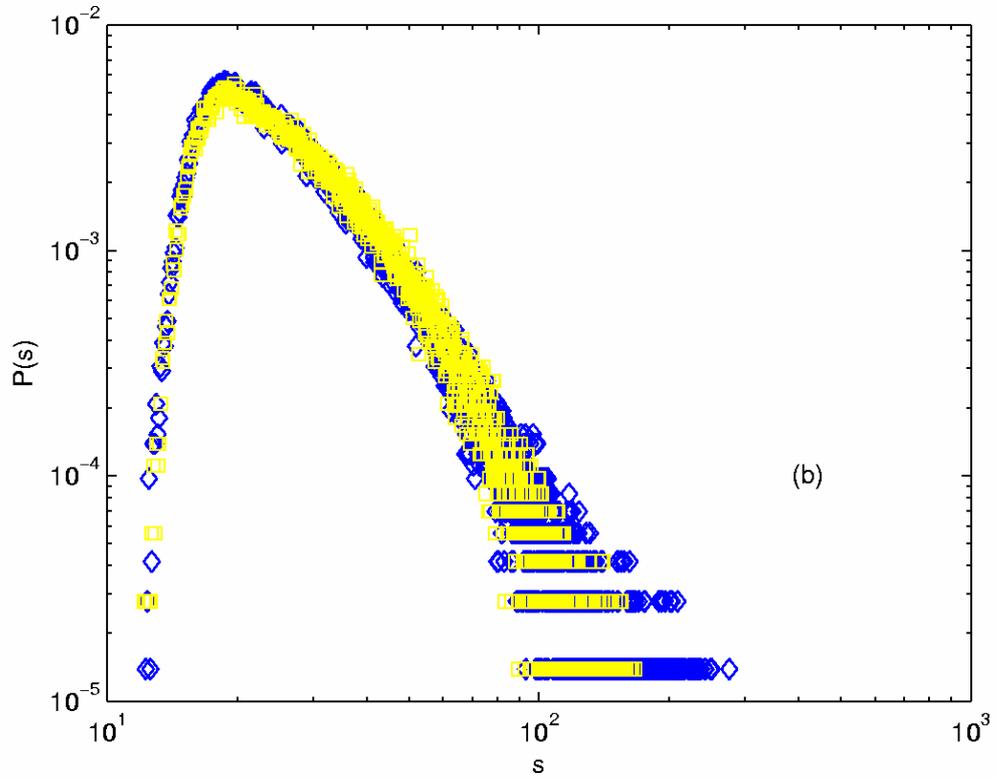

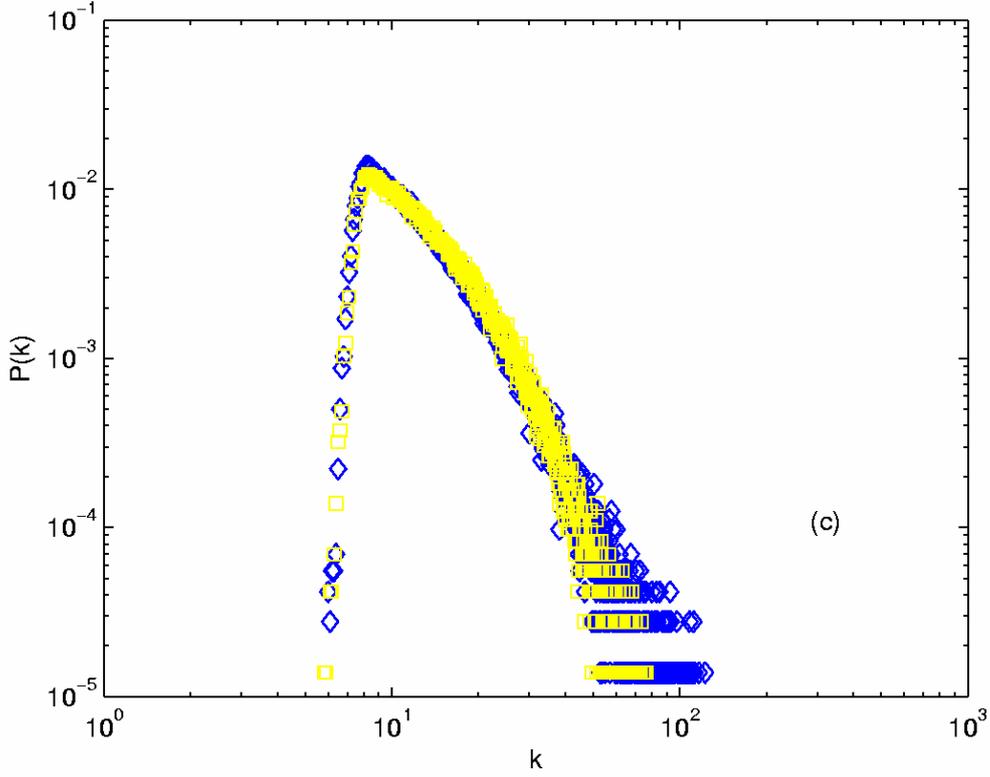

FIG. 3 Numerical results in which $b=2$, $n=20$, $m=10$ and $\zeta=2.5$. (a) Distribution $P(k)$ of total degree. The symbols correspond to different value of $p$, i.e., $p=1$ ($\diamond$), $20$ ($\square$). (b) Distribution $P(w)$ of strength. (c) Distribution $P(k)$ of the first subgraph. The data are averaged over $10$ independent runs of size $N=72000$.

Further more, for mimicking real systems. We assume that for $l \leq p$, $d_l = l$, but when $l > p$, $d_l$ drops linearly from $p$ to unity, namely, $d_l - p = (1-p)(l-p)/(n-p)$, where $p \in [1, n]$ is a integer and $n$ is the number of subgraphs we divide the network into. This assumption indicates that when the edge weight is not too large, the contribution of an edge increases as its weight increases. But when the edge weight is over a threshold, which is governed by parameter $p$, the more weight the edge is, the less it contributes. For instance, when $p=1$, all edges contribute the same, this is the case in degree preferential attachment. When $p=n$, the contribution is proportional to the edge weight all the time, so we turn to the weight preferential attachment. Then, we assume $c_j$ has a power-law form $c_j = j^{-\zeta}$, where $\zeta$ is a positive number. That is, the likelihood that the new edge weight is $\overline{w}_j$ decreases rapidly as $\overline{w}_j$ increases. Yet, in this paper, we will study the simplest case of the initial attractiveness $\mathbf{b} = (b_1, b_2, \cdots, b_n)^T$, in which all elements have the same value, $b_j = b$, for $j = 1, 2, \cdots, n$. When $b=0$, equation (16) gives $\gamma = 3$, which is independent from $p$, $\zeta$, and $m$. That means when there is no initial attractiveness, no matter we use the degree preferential attachment ($p=1$), or the weight preferential attachment ($p=n$), or others, the model yields power law distributions of degree (including total degree and degree in each subgraph) and strength with exponents equals exactly to three. This result is confirmed by numerical simulations, as seen in Fig. 2. When $b \neq 0$, the model also yields power law distributions for large $k$ and $s$. But this time, just as equation (16) indicates, the exponents depend on what kind of preferential attachment we choose. This is also observed in numerical results, see Fig.3. On the other hand, according to equation (19), since we assume $c_j$ has a power-law form $c_j = j^{-\zeta}$, the distribution of single edge weight $P(w)$ will also follows a power-law form $P(w) \sim w^{-\zeta}$ with the same exponent as that of $c_j$, and independent from $p$, $m$ and $b$. This result is confirmed by numerical simulations too, as seen in Fig. 4.

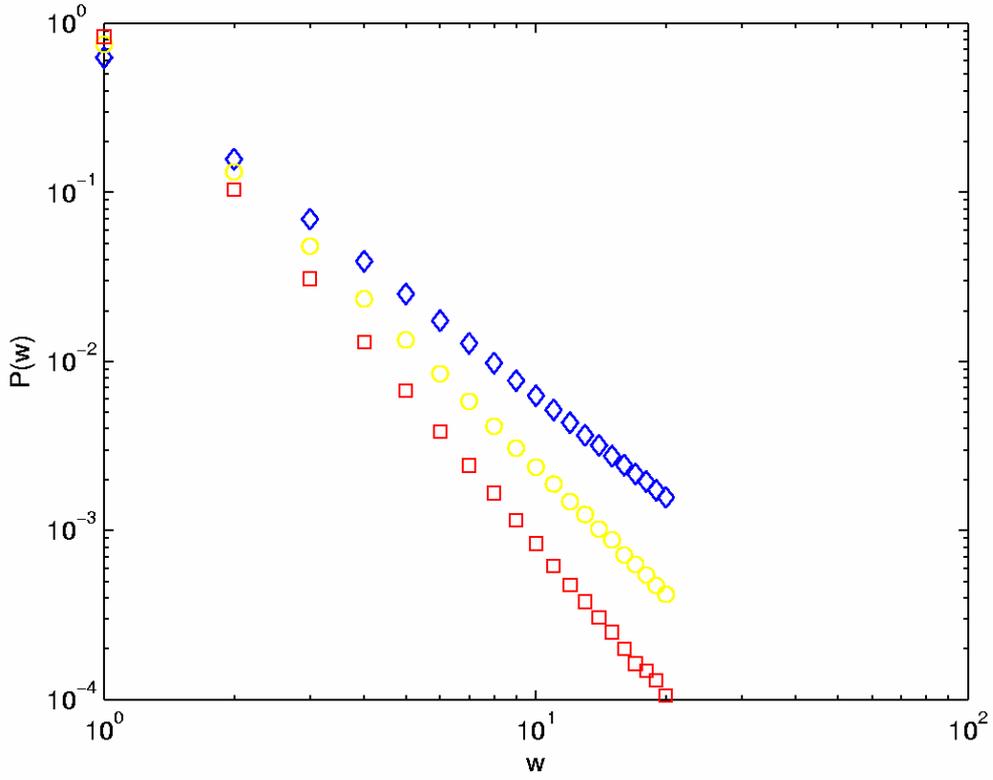

FIG. 4 (Color online). Distribution $P(w)$ of single edge weight with $b = 2$, $n = 20$, $m = 10$, $p = 5$. The symbols correspond to different value of $\zeta$, i.e., $\zeta = 2$ ($\diamond$), $2.5$ ($\circ$), $3$ ($\square$). The data are averaged over 10 independent runs of size $N = 72000$.

In conclusion, the division of edge-weight range discretizes one weighted network into several unweighted subgraphs. And with a proper definition of equivalence of two vertices, we can get one preferential attachment which accurately reflects the contribution of each subgraph. The generalized preferential attachment can tell us not only the chance that each already-existing vertex being connected but also how much weight each new edge has. And by using this preferential attachment we can generate a network which displays power-law distributions of degree, strength, and single edge weight.

Thanks Prof. Z.B. Li, C.S. He, Z.F. Chen, and W. Pang for fruitful discussions.